\documentclass[12pt,amsmath,amssymb,nofootinbib]{revtex4}
\usepackage{amsmath,amsfonts,amsthm,amscd,amssymb,latexsym}
\usepackage{bm}
\usepackage{dcolumn}
\usepackage{graphicx}
\usepackage{epstopdf}
\usepackage{color}
\usepackage{epsf}
\usepackage{epsfig}
\usepackage{graphicx, epic, eepic, color}
\usepackage{soul}

\begin{document}

\title{Properties and Patterns of Polarized Gravitational Waves}

\author{Bahram \surname{Mashhoon}$^{1,2}$} 
\email{mashhoonb@missouri.edu}
\author{Sohrab \surname{Rahvar}$^{3}$}
\email{rahvar@sharif.edu}

\affiliation{
$^1$School of Astronomy,
Institute for Research in Fundamental Sciences (IPM),
Tehran 19395-5531, Iran\\
$^2$Department of Physics and Astronomy,
University of Missouri, Columbia,
Missouri 65211, USA\\
$^3$Department of Physics, Sharif University of Technology, Tehran 11365-9161, Iran\\}

\date{\today}
\begin{abstract}
We discuss polarization of gravitational radiation within the standard framework of linearized general relativity. The recent experimental discovery of gravitational waves provides the impetus to revisit the implications  of the spin-rotation-gravity coupling for polarized gravitational radiation; therefore, we consider the coupling of helicity of gravitational waves to the rotation of an observer or the gravitomagnetic field of a rotating astronomical source.  Observational possibilities regarding polarization-dependent effects in connection with future gravitational wave detectors are briefly explored. 
\end{abstract}

\maketitle

\section{Introduction}

Gravitational radiation is a natural component of a field theory of the gravitational interaction. In general relativity (GR), the gravitational field is represented by spacetime curvature. Thus, perturbations in the Riemannian curvature of spacetime are expected to propagate as gravitational waves at the speed of light. Gravitation is the weakest of the known forces of nature; nevertheless, gravitational radiation could provide an important new window to observe high-energy processes in the universe. The generation, propagation and detection of gravitational radiation in GR have been extensively studied~\cite{MTW}.

In linearized general relativity,  gravitational waves constitute a small symmetric tensor perturbation of the background spacetime metric. To simplify matters, let us consider free gravitational waves on Minkowski spacetime background; therefore, we have a slight perturbation of a background global inertial frame with Cartesian coordinates $x^\mu = (ct, \mathbf{x})$ in Minkowski spacetime.  The perturbed spacetime has metric tensor $\eta_{\mu \nu} + h_{\mu \nu}(x)$, where $\eta_{\mu \nu} = \rm{diag}(-1, 1, 1, 1)$ is the Minkowski metric tensor and $h_{\mu \nu}$ transforms as a tensor under the Poincar\'e group.  In our convention, Greek indices run from 0 to 3, while Latin indices run from 1 to 3; moreover,  we use units such that $c= 1$, unless specified otherwise. 

A slight change in the background spacetime coordinates induces a gauge transformation in $h_{\mu \nu}$. With a proper transverse gauge condition for the trace-reversed form of the linear perturbation, the latter satisfies the standard wave equation in Minkowski spacetime~\cite{MTW}. Therefore, a plane gravitational wave of frequency $\omega$ and wave vector $\mathbf{k}$, $\omega = c |\mathbf{k}|$, has an   amplitude $h_{\mu \nu} \propto \exp(i\eta_{\alpha \beta}k^\alpha x^\beta)$, where $k^\mu = (\omega/c, \mathbf{k})$ is the null propagation 4-vector. Our treatment is linear throughout; therefore, we can use complex amplitudes with the proviso that only the real part has physical significance.  

In principle, gravitational wave properties are measured by the fundamental observers at rest in spacetime; indeed, such an  observer carries an orthonormal tetrad frame $e^{\mu}{}_{\hat \alpha}$, which corresponds to $\delta^\mu_\alpha$ in the absence of the perturbation. Here, hatted indices enumerate tetrad axes in the local tangent space, while indices without hats are regular spacetime indices.  Furthermore, we employ the standard TT gauge~\cite{MTW}, where $h_{0\mu} = 0$ and $h_{ij}$ is transverse and traceless, namely, $h^{ij}{}_{,j} = 0$ and $h^{i}{}_{i} = 0$. The spacetime metric is then $dS^2 = -dt^2 + (\delta_{ij}+h_{ij})\,dx^idx^j$. 

In 2015, earth-based laboratories LIGO and Virgo succeeded in detecting gravitational waves~\cite{LIGOScientific:2016aoc}. A transient gravitational signal with frequency in the range 35 Hz to 250 Hz was originally observed with a peak amplitude of about $10^{-21}$. The detection of such signals has continued and the results are in agreement with general relativity theory~\cite{Akhshi:2021nsy}. The signals have been interpreted to originate from the merger of gravitationally collapsed astronomical systems. The LIGO-Virgo collaboration has thus far detected short-duration transients that last from about a millisecond to about a second. Such transient signals are the main focus of the present work.

The main purpose of this paper is to describe certain polarization-dependent effects of gravitational waves (GWs) and explore briefly the possibility of measuring them via GW detectors. In the rest of this introductory section, we briefly describe the polarization of linearized GWs; moreover, we note that the standard Doppler effect and aberration, as measured in effect by background inertial observers moving uniformly with velocity $\mathbf{v}$, are independent of the polarization of the wave. On the other hand, as described in Section II, GW polarization needs to be taken into account for rotating observers due to the coupling of the helicity of GWs to the rotation of the observer's spatial frame. The treatment of helicity-rotation coupling is extended to helicity-gravity coupling in Sections III and IV by means of the gravitational Larmor theorem. That is, we investigate the coupling of helicity of GWs to the gravitomagnetic field of a rotating astronomical source. Section V is devoted to the gravitational Faraday rotation of linearly polarized GWs in the gravitomagnetic field of a rotating mass. In Section VI, the corresponding gravitational time delay of circularly polarized GWs is calculated. We discuss various observational possibilities in connection with transient signals. Finally, Section VII contains a brief discussion of our results. 

\subsection{Polarization of GWs}

For our present purposes, we assume that the plane gravitational wave propagates along the positive $z$ direction in the background global inertial frame. Then, $h_{ij} \propto \exp[-i\omega (t-z)]$ and transversality implies that we are left with two independent states of gravitational radiation given by $h_{11} = -h_{22}$ and $h_{12} = h_{21}$. We can write $(h_{ij}) = h_{11}(t- z)\, \Pi_1 + h_{12}(t- z)\, \Pi_2$, where 
\begin{equation}\label{1}
\Pi_1 = 
\begin{bmatrix}
1&0&0 \\
0&-1&0 \\
0&0&0 
\end{bmatrix}
\,, \qquad \Pi_2 =
\begin{bmatrix}
0&1&0 \\
1&0&0 \\
0&0&0 
\end{bmatrix}
\,
\end{equation} 
are $3\times3$ traceless matrices known as the ``plus" $(\oplus)$  and ``cross" $(\otimes)$ \emph{linear} polarization states of the incident radiation, respectively. These matrices represent the two \emph{independent} linear polarization states of the gravitational wave and are such that $\rm{tr}(\Pi_1^2) = \rm{tr}(\Pi_2^2) = 2$ and $\rm{tr}(\Pi_1\Pi_2) = \rm{tr}(\Pi_2\Pi_1) = 0$.

Next, let us consider a rotation of the Cartesian coordinates by an angle $\vartheta$ about the direction of propagation of the wave; that is, $z' = z$ and
\begin{equation}\label{2}
x' = x \cos \vartheta + y \sin\vartheta\,, \qquad y' = - x \sin \vartheta + y \cos\vartheta\,.
\end{equation}
Under this rotation, the spacetime metric remains invariant, namely, 
\begin{equation}\label{3}
dS^2 = -dt^2 + (\delta_{ij}+h'_{ij})dx'^idx'^j\,;
\end{equation}
hence, 
\begin{equation}\label{4}
h'_{11} = h_{11} \cos (2\vartheta) + h_{12} \sin(2\vartheta)\,, \qquad h'_{12} = - h_{11} \sin(2 \vartheta) + h_{12} \cos(2\vartheta)\,.
\end{equation}
Alternatively, we can write $(h'_{ij}) = h_{11}(t, z)\, \Pi'_1 + h_{12}(t, z)\, \Pi'_2$, where
\begin{equation}\label{5}
\Pi'_{1} = \Pi_{1} \cos (2\vartheta) - \Pi_{2} \sin(2\vartheta)\,, \qquad \Pi'_{2} =  \Pi_{1} \sin(2 \vartheta) + \Pi_{2} \cos(2\vartheta)\,.
\end{equation}
In any case, with $2\vartheta = \pi/2$, we find $(h'_{11}, h'_{12}) = ( h_{12}, - h_{11})$ or, alternatively, $(\Pi'_{1}, \Pi'_{2}) = (- \Pi_{2}, \Pi_{1})$; that is, a rotation by an angle of $\vartheta = \pi/(2s)$, with $s=2$ in the gravitational case, turns one \emph{independent} linear polarization state into another. 

We define the circular polarization states of gravitational radiation in complete analogy with electrodynamics, namely, 
\begin{equation}\label{6}
\Pi_{\pm} = \frac{1}{\sqrt{2}} (\Pi_{1} \pm i \Pi_{2})\,.
\end{equation}
In the case under consideration, $(h_{ij}) = C_{+}(t, z)\, \Pi_{+} + C_{-}(t, z)\, \Pi_{-}$, where 
\begin{equation}\label{7}
C_{\pm} = \frac{1}{\sqrt{2}} (h_{11} \mp i h_{12})\,.
\end{equation}
Under the rotation of angle $\vartheta$ about the direction of wave propagation, we find
\begin{equation}\label{8}
C'_{+} = e^{2i\vartheta} C_{+}\,, \qquad C'_{-} = e^{-2i\vartheta} C_{-}\,,
\end{equation}
 consistent with the interpretation of $C_{+}$ and $C_{-}$  as the amplitudes of the positive and negative helicity states of gravitational radiation ($s=2$), respectively. 
 
In view of our linear perturbation scheme, it is natural to write the perturbation amplitude due to an incident plane wave packet in the form
\begin{equation}\label{9}
h_{ij} = \rm{Re}\sum_{\omega} [\hat{C}_{+}(\omega)\, \Pi_{+} + \hat{C}_{-}(\omega)\, \Pi_{-}]\,e^{-i \omega(t-z)}\,.
\end{equation}
For an incident wave of frequency $\omega$, the gravitational energy flux, averaged over the wave period, along the direction of the wave propagation is given by
\begin{equation}\label{10}
\frac{\omega^2}{32 \pi} \left(|\hat{C}_{+}(\omega)|^2 + |\hat{C}_{-}(\omega)|^2\right)\,.
\end{equation}

The formal aspects of the polarization properties of linearized gravitational waves have been treated in~\cite{AnBr, Mash78}; in particular, a detailed description of the Stokes parameters and the corresponding Poincar\'e sphere for partially polarized gravitational radiation is contained in Appendix A of Ref.~\cite{Mash78}. 

Suppose the spacetime metric is of the form $-dt^2 + g_{ij}(x) dx^idx^j$; then, there is a theorem in  general relativity which states that test particles that are spatially at rest in this spacetime follow geodesics and are therefore free. In the gravitational wave spacetime under consideration here, $g_{ij} = \delta_{ij} + h_{ij}(x)$. We can therefore choose a static test particle as our reference observer and establish a quasi-inertial Fermi normal coordinate system along its geodesic world line. We can then study the influence of the curvature-based gravitoelectric and gravitomagnetic fields associated with the gravitational wave on physical systems. This approach to GW effects has been treated in detail in Refs.~\cite{Mashhoon:2021qtc, Ruggiero:2020oxo, Ruggiero:2021qnu, Ruggiero:2022gzl} and the references cited therein. 

\subsection{Doppler Effect and Aberration for GWs}

We have thus far dealt with observers that have remained spatially at rest. Let us now imagine that these observers all move with velocity $\mathbf{v}$. What are the GW properties as measured by these boosted observers? Their motion is affected by the presence of the wave; however, to calculate the corresponding wave amplitude, boosted observers can be considered to be uniformly moving and ``inertial" in the linear perturbation approach under consideration in this paper.  For inertial observers in the background Minkowski spacetime, $h_{ij}$ transforms as a tensor under Lorentz transformations. This affects the amplitude of the wave, but its phase remains invariant. It follows in general from the invariance of the phase of the wave $\Phi = \eta_{\alpha \beta} k^\alpha x^\beta = -\omega t + \mathbf{k}\cdot \mathbf{x}$ under Lorentz transformations that the wave frequency $\tilde{\omega}$ and wave vector $\tilde{\mathbf{k}}$ as measured by essentially uniformly moving observers with unperturbed 4-velocity $u^\mu = \gamma (1, \mathbf{v})$, where $\gamma$ is the Lorentz factor, are given by
\begin{eqnarray}\label{a}
\tilde{ \omega} & = & -k_{\alpha}u^\alpha = \gamma (\omega -\mathbf{v}\cdot \mathbf{k})\,,\\
\label{b} \tilde{\mathbf{k}} &=&\mathbf{k}+\frac{\gamma -1}{v^2}(\mathbf{v}\cdot \mathbf{k})\mathbf{v}-\frac{1}{c^2}\,\gamma\, \omega\, \mathbf{v}\,.
\end{eqnarray}
That is, $k^\mu$ and $\tilde{k}^\mu$ are simply related by a pure boost, which is the standard result  in the treatment of the properties of electromagnetic waves as well. For light, the first relation~\eqref{a} is the relativistic expression for the Doppler effect and the second relation~\eqref{b} expresses the aberration of starlight whose original discovery by James Bradley~\cite{Brad} eventually led to the development of Lorentz transformations as the relativistic generalization of the Galilean transformations. 

Let us note that we can write
\begin{equation}\label{c}
\frac{\tilde{\mathbf{k}} - \mathbf{k}}{|\mathbf{k}|} =  - \frac{\mathbf{v}}{c}\,\left[1 + (\gamma -1)\left(1- \frac{c\,\mathbf{v}\cdot \mathbf{k}}{v^2 k}\right)\right]\,,
\end{equation}
from which we recover the magnitude of the traditional aberration angle $\alpha \approx v/c$. For ``inertial" observers, the polarization of the incident linearized gravitational radiation makes no contribution to the relativistic expressions for the aberration and Doppler effect.  As described in the next section,  the situation is different, however, for rotating observers. 

\section{Reception of Radiation by Accelerated Observers}

In contrast to ideal inertial observers,  realistic observers in Minkowski spacetime are all more or less accelerated.  An accelerated world line in the background global inertial frame has translational acceleration $a^\mu = d^2 x^\mu/d\tau^2$, where $\tau$ is the proper time of the observer along its path. The accelerated observer in the background global inertial frame carries its adapted orthonormal tetrad frame $e^{\mu}{}_{\hat \alpha}(\tau)$ such that 
\begin{equation}\label{A}
\frac{d e^{\mu}{}_{\hat \alpha}}{d\tau} = \mathbb{A}_{\hat \alpha}{}^{\hat \beta} \,e^{\mu}{}_{\hat \beta}\,,
\end{equation}
where $\mathbb{A}_{\hat \alpha \hat \beta}$ is the observer's acceleration tensor. It is antisymmetric,  $\mathbb{A}_{\hat \alpha \hat \beta} = - \mathbb{A}_{\hat \beta \hat \alpha}$,  due to the tetrad orthonormality condition, namely, 
\begin{equation}\label{B}
\eta_{\mu \nu} \,e^\mu{}_{\hat \alpha}\,e^\nu{}_{\hat \beta}= \eta_{\hat \alpha \hat \beta}\,.
\end{equation}

The acceleration tensor is analogous to the Faraday tensor of electrodynamics and can be naturally decomposed into its ``electric" and ``magnetic" parts, namely, $\mathbb{A}_{\hat \alpha \hat \beta} \mapsto (-\mathbf{g}, \boldsymbol{\Omega})$, where $\mathbf{g}(\tau)$ and $\boldsymbol{\Omega}(\tau)$ are spacetime scalars  that represent the translational and rotational accelerations  of the  observer, respectively. That is, $g_{\hat i} = \mathbb{A}_{\hat 0 \hat i} = a_\mu e^{\mu}{}_{\hat i}$ specifies the deviation of observer's world line  from a geodesic, while $\boldsymbol{\Omega}$, $\mathbb{A}_{\hat i \hat j} = \epsilon_{\hat i \hat j \hat k}  \Omega^{\hat k}$, is the angular velocity of the rotation of the observer's spatial frame with respect to a locally nonrotating (i.e., Fermi-Walker transported) frame. 

Imagine a test electromagnetic radiation field $F_{\mu \nu}(x)$ in Minkowski spacetime and a class of accelerated observers with adapted tetrad frame field $\chi^{\mu}{}_{\hat \alpha}(x)$. The radiation field as measured by the accelerated observers is given by the projection of the Faraday tensor on the tetrad frame field, namely, $F_{\mu \nu}\chi^{\mu}{}_{\hat \alpha}\chi^{\nu}{}_{\hat \beta} = F_{\hat \alpha \hat \beta}(x)$. Upon Fourier transformation of this scalar, one can, in principle, determine the measured frequency and wave vector content of the radiation field.  To illustrate this idea, let us consider the class of static observers in the background global inertial frame  that refer their measurements to axes that rotate uniformly with positive frequency  $\Omega$ about the $z$ axis. Plane electromagnetic waves of frequency $\omega \gg \Omega$ and definite helicity propagate along the positive $z$ axis. In the case of positive (negative) helicity waves, the electric and magnetic fields rotate in the positive (negative) sense with frequency $\omega$ about the direction of propagation of the wave.  As seen by the rotating static observers, the electric and magnetic fields of the incident radiation with positive (negative) helicity, rotate with frequency $\omega - \Omega$ ($\omega + \Omega$) in the positive (negative) sense about the $z$ axis. Indeed, Fourier analysis in this case reveals that~\cite{BMr, Hauck:2003gy}
\begin{equation}\label{Ba}
\omega' = \omega \mp \Omega\,, \qquad  k_{z}' = k_z\,.
\end{equation}
This example demonstrates the coupling of the helicity of the electromagnetic radiation with the rotation of the observer.  On the other hand, the transverse Doppler effect~\eqref{a} implies $\tilde{\omega} = \omega$. It is clear that the relative deviation from the transverse Doppler effect is $\mp\, \Omega/\omega$, which vanishes in the high-frequency limit (i.e., $\omega \to \infty$). That is, when the wavelength of the radiation $\lambda = 2\pi c/\omega$ is negligibly small compared to all other relevant lengths in the problem ( i.e., $c/\Omega$ in the present case), wave effects can be neglected. 

A similar analysis can be carried out for linearized gravitational radiation~\cite{Ramos:2006sb}. In this case, the gravitational field is represented by the Riemann tensor $R_{\mu \nu \rho \sigma}$. Therefore, the field as measured by the accelerated observers is given in our linear perturbation scheme by the projection of the Riemann tensor on the observer's orthonormal tetrad frame field, $R_{\mu \nu \rho \sigma}\chi^{\mu}{}_{\hat \alpha}\chi^{\nu}{}_{\hat \beta}\chi^{\rho}{}_{\hat \gamma}\chi^{\sigma}{}_{\hat \delta} = R_{\hat \alpha \hat \beta \hat \gamma \hat \delta}$. This scalar quantity can be expressed as a $6\times 6$ matrix via the symmetries of the Riemann curvature tensor,
\begin{equation}\label{C} 
\begin{bmatrix}
\mathbb{E}&\mathbb{B} \\
\mathbb{B}&-\mathbb{E} 
\end{bmatrix}\,,
\end{equation} 
where the matrix indices run over the set $\{01, 02, 03, 23, 31, 12\}$. This is the general form of the measured curvature in the Ricci-flat case, where the Riemann tensor degenerates into the Weyl conformal curvature tensor. Here, $\mathbb{E}$ and $\mathbb{B}$ are  symmetric and traceless $3\times 3$ matrices that represent the gravitoelectric and gravitomagnetic fields of the incident gravitational radiation, respectively. 

To illustrate these ideas, we note that for plane gravitational waves of frequency $\omega$ propagating along the $z$ axis,  the curvature as measured by the static inertial observers in the background Minkowski spacetime is given by
\begin{equation}\label{D}
\mathbb{E} = \frac{1}{2} \omega^2 h_{ij} = \frac{1}{2} \omega^2
\begin{bmatrix}
h_{11}&h_{12}&0 \\
h_{12}&-h_{11}&0 \\
0&0&0 
\end{bmatrix}
\,, \qquad \mathbb{B} =  \frac{1}{2} \omega^2
\begin{bmatrix}
h_{12}&-h_{11}&0 \\
-h_{11}& -h_{12}&0 \\
0&0&0 
\end{bmatrix}
\,.
\end{equation} 
Moreover, $\mathbb{E} \mp i \mathbb{B} = \omega^2 C_{\pm} \Pi_{\pm}$, which is analogous to electrodynamics in connection with electromagnetic radiation of definite helicity. Indeed, the Riemann-Silberstein vectors $\mathbf{E} + i \mathbf{B}$  and $\mathbf{E} - i \mathbf{B}$ characterize complex electromagnetic fields that represent positive and negative helicity waves, respectively. 

Let us now consider the static \emph{rotating} observers and their measurements of the properties of the plane gravitational wave of frequency $\omega$ propagating along the positive $z$ direction in the background global inertial frame. In this case, a detailed examination reveals that the measured frequency and wave vector of the GW are given by~\cite{Ramos:2006sb}
\begin{equation}\label{Da}
\omega' = \omega \mp 2\, \Omega\,, \qquad  k_{z}' = k_z\,,
\end{equation}
in agreement with the spin-2 nature of linearized gravitational radiation.  

It is possible to see that our helicity-rotation coupling result for gravitational waves~\eqref{Da} is in conformity with our description of the polarization properties of gravitational waves. Indeed, with  $\vartheta = \Omega t$,  Equation~\eqref{8} implies 
\begin{equation}\label{Db}
C'_{\pm} = e^{\pm2i\Omega t}\,C_{\pm} \propto e^{-i(\omega \mp 2\Omega)t + i \omega z}\,.
\end{equation}

Thus far the direction of incidence of the gravitational wave has been the same as the axis of rotation of the observer. We must now consider the case of oblique incidence. 

\subsection{Oblique Incidence}

The case of oblique incidence is in general more complicated, even in the simple example of the rotating static observers. Working to first order in $\Omega/\omega$ within the quasi-classical approximation, it can be shown that 
\begin{equation}\label{E}
\omega' = \omega - \mathbf{H}\cdot \boldsymbol{\Omega}\,,\qquad \mathbf{k}' = \mathbf{k}\,,
\end{equation}
where $\mathbf{H}$ is the helicity vector of the radiation,
\begin{equation}\label{F}
\mathbf{H} = \pm\, s\,\hat{\mathbf{k}}\,
\end{equation}
and $\hat{\mathbf{k}} = \mathbf{k}/|\mathbf{k}|$ is the unit propagation vector. Here, $s=1$ for the photon~\cite{BMr, Hauck:2003gy} and $s=2$ for the graviton~\cite{Ramos:2006sb}.  We emphasize that the existence of 
$- \mathbf{H}\cdot \boldsymbol{\Omega}$ term in Equation~\eqref{E} goes beyond the hypothesis of locality of the standard theory of relativity,  which states that an accelerated observer at each instant along its world line is  equivalent to an otherwise identical momentarily comoving inertial observer. In effect, locality requires that  we apply Lorentz transformations point by point along the world line of the accelerated observer.

The spin-rotation coupling is a manifestation of the inertia of intrinsic spin. A comprehensive summary of spin-rotation-gravity coupling is contained in Ref.~\cite{BMB}. For more recent observational results involving neutrons, see Ref.~\cite{Danner}. 

The corresponding generalizations of relations~\eqref{a} and~\eqref{b} for arbitrary accelerated observers can be obtained straightforwardly from Equation~\eqref{E} by an instantaneous pure boost in accordance with the hypothesis of locality. The resulting modified expressions for the Doppler effect and aberration are given  in the quasi-classical approximation by
\begin{eqnarray}
\label{G} \omega' &=&\gamma [(\omega - \mathbf{H}\cdot \boldsymbol{\Omega})-\mathbf{v}\cdot \mathbf{k}]\,,\\
\label{H}\mathbf{k}' &=&\mathbf{k}+\frac{\gamma -1}{v^2} (\mathbf{v}\cdot \mathbf{k})\mathbf{v} - \frac{1}{c^2}\,\gamma (\omega - \mathbf{H} \cdot \boldsymbol{\Omega})\mathbf{v}\,.
\end{eqnarray}
In these formulas, $\mathbf{v}$ and $\boldsymbol{\Omega}$ depend in general upon time. In the eikonal limit ($\lambda \to 0$, $\omega \to \infty$) of Equations~\eqref{G} and~\eqref{H}, the $\mathbf{H}\cdot \boldsymbol{\Omega}$ term disappears and we recover the hypothesis of locality in which the phase of the wave is  invariant under Lorentz transformations.  Moreover, a comment is in order here regarding the fact that the spin of the incident radiation field couples to the rotation of the observer but not to its translational acceleration. In this connection, we note that the spin-rotation coupling term $- \gamma \,\mathbf{H}\cdot \boldsymbol{\Omega}$ is consistent with invariance under parity and time reversal, while the corresponding coupling involving translational acceleration violates both parity and time reversal invariance. Besides, various investigations have not revealed a corresponding spin-acceleration-gravity coupling~\cite{BMa, Mashhoon:2000jq, Bini:2004kz, Bini:2004ay, Mashhoon:2013jaa}.

\subsection{Rotating Detector of GWs}

To illustrate helicity-rotation contributions to the Doppler effect and aberration, we write Equations~\eqref{G} and~\eqref{H} in the form
\begin{equation}\label{H1}
\omega' \approx \tilde{\omega} - \gamma \mathbf{H}\cdot \boldsymbol{\Omega}\,
\end{equation}
and
\begin{equation}\label{H2}
 \mathbf{k}' \approx  \tilde{\mathbf{k}} + \gamma \frac{\mathbf{v}}{c^2}\mathbf{H}\cdot \boldsymbol{\Omega}\,.
\end{equation}
 The equation for aberration now takes the form
\begin{equation}\label{H3}
\frac{\mathbf{k}' - \mathbf{k}}{|\mathbf{k}|} =  - \frac{\mathbf{v}}{c}\left[1 + (\gamma -1)\left(1- \frac{c\,\mathbf{v}\cdot \hat{\mathbf{k}}}{v^2}\right)\right] + \gamma \frac{\mathbf{v}}{c}\frac{\mathbf{H}\cdot \boldsymbol{\Omega}}{\omega}\,.
\end{equation}

Suppose an observer is on a satellite that moves with speed $v$. There is the standard effect of aberration of starlight such that the direction of a source is displaced by the aberration angle $\alpha$ of about $v/c$. If the satellite rotates (for attitude control and stability, say) with frequency $\Omega$ and the incident radiation of frequency $\omega$ is circularly polarized, then positive helicity waves and negative helicity waves give slightly different directions for the source. These occur symmetrically about the main aberration-corrected direction of the source with an angle 
\begin{equation}\label{J}
\alpha'_{\pm} \approx s \frac{v}{c}\frac{\Omega}{\omega}\cos \xi\,, 
\end{equation}
where $\xi$ is the angle formed by the direction of incidence of the wave and the axis of rotation of the observer,  $s = 1$ for electromagnetic radiation and $s=2$ for gravitational radiation. This effect could be interesting for rotating detectors of gravitational radiation. In this connection, we note that visible light has frequencies $\nu = \omega/(2\pi)$ of order $10^{14}$ Hz, while gravitational waves that are currently being detected have frequencies  of about 100 Hz.

To illustrate helicity-rotation coupling for detectors such as LIGO and Virgo that are fixed on the rotating earth,  we focus here on the $- \mathbf{H}\cdot \boldsymbol{\Omega}$ term in Equation~\eqref{H1} and ignore for the moment any Doppler  frequency shift due to the rotation velocity of the detectors.  We imagine that these detectors can perform accurate measurement of the frequency of incident gravitational wave. Moreover, we assume the earth has radius $R_{\oplus}$ and ignore its deviations from spherical symmetry. The angular velocity of a  detector $\mathbb{D}$ with respect to the  background global inertial frame is given by the component of $\boldsymbol{\Omega}_{\oplus}$ along the local vertical direction on the rotating earth and can be written as 
\begin{equation}\label{J1}
\boldsymbol{\Omega}_{\mathbb{D}} = \Omega_{\oplus} \left[ \mathbf{e}_x \cos \ell \cos(\Omega_{\oplus} t + \varphi_0) + \mathbf{e}_y \cos \ell \sin(\Omega_{\oplus} t + \varphi_0) + \mathbf{e}_z \sin \ell \right]\,, 
\end{equation}
where ${\Omega}_{\oplus}$ is the magnitude of the angular velocity of the earth along its rotation axis $\mathbf{e}_z$, $\ell$ is the geographic latitude of the location of the gravitational wave detector $\mathbb{D}$ measured from the equator and $\varphi_0$ represents its geographic longitude. Here, $(\mathbf{e}_x,  \mathbf{e}_y,  \mathbf{e}_z)$ are unit vectors along the fixed  spatial axes of the background Cartesian coordinate system whose origin is at the center of the earth that is assumed to be at rest. For a gravitational wave incident on detector $\mathbb{D}$ along the direction of polar angles $(\theta,\phi)$ with respect to the background Cartesian coordinate system, we have 
\begin{equation}\label{J2}
\mathbf{H} = \pm 2\,\hat{\mathbf{k}}\,, \qquad \hat{\mathbf{k}} = \mathbf{e}_x \sin \theta \cos \phi + \mathbf{e}_y \sin \theta \sin \phi + \mathbf{e}_z \cos \theta\,. 
\end{equation}
Therefore,  we expect the measured relative frequency change of  the gravitational waves due to helicity-rotation coupling would be 
\begin{align}\label{J3}
\nonumber  \frac{\omega - \omega_{\mathbb{D}}}{\omega} \approx {}& \pm 2 \frac{\Omega_{\oplus}}{\omega}\,[\sin \theta \cos \phi \cos \ell \cos(\Omega_{\oplus} t +\varphi_0) \\
 {}&  + \sin \theta \sin \phi \cos \ell \sin(\Omega_{\oplus} t + \varphi_0)+ \cos \theta \sin \ell]\,,
\end{align}
where the plus and minus signs represent the two possible circular polarization states of the gravitational  wave.
For gravitational waves with frequency of about $100$ Hz, $2\Omega_{\oplus}/\omega$ would be about $2.3 \times 10^{-7}$, since for the earth $\Omega_{\oplus}/(2\pi) \approx 1.16 \times 10^{-5}$ Hz. Assuming a series of ideal gravitational wave detectors that can measure the frequency of the signal with high accuracy, we want to investigate the parameters that detectors can determine via the frequency shift.

The frequency of a gravitational wave event measured in detector $\mathbb{D}$ is given by 
\begin{equation}\label{J4}
\omega_{\mathbb{D}} \approx \omega - \mathbf{H}\cdot \boldsymbol{\Omega}_{\mathbb{D}} - \frac{\omega}{c}\,\hat{\mathbf{k}} \cdot \mathbf{v}_{\mathbb{D}}\,,
\end{equation}
where 
\begin{equation}\label{J5}
 \mathbf{v}_{\mathbb{D}} = R_{\oplus} \Omega_{\oplus} \cos \ell \left[- \mathbf{e}_x  \sin (\Omega_{\oplus} t + \varphi_0) + \mathbf{e}_y \cos (\Omega_{\oplus} t + \varphi_0)\right]\,,
\end{equation}
$R_{\oplus} \Omega_{\oplus}/c$ is about $1.5 \times 10^{-6}$  and we have neglected  $(v_{\mathbb{D}}/c)^2$ terms for the sake of simplicity.  For gravitational waves with $\nu \approx 100$ Hz, the relative strength of the helicity-rotation shift in Eq.~\eqref{J4}, $2\Omega_{\oplus}/\omega \approx 2.3 \times 10^{-7}$, is only about an order of magnitude smaller than the relative strength of the Doppler shift $R_{\oplus} \Omega_{\oplus}/c \approx 1.5 \times 10^{-6}$.
Measuring the left-hand side of Equation~\eqref{J4} via detector $\mathbb{D}$, the right-hand side is a function of the original frequency of the source  $\omega$ as well as $\theta$ and $\phi$ that specify the direction of incidence of the gravitational wave. Therefore,  we need at least three detectors at fixed longitudes and latitudes to measure the frequency of the gravitational wave and its direction of incidence~\cite{GWarray, SathyaSchutz:LivingReviews}. We note that in this method the frequency shifts due to the motions of the local frames of the detectors provide us with the angular position of the source on the sky. 

It is worthwhile to mention that the situation would be different if we employ continuous gravitational waves instead of short pulses. For continuous waves from a pulsar, for instance, we can use a single detector instead of three and make measurements at three different times or more within 24 hours of exposure. The resulting different values of the phase $\varphi = \Omega_{\oplus} t + \varphi_0$ in Equations~\eqref{J3} and~\eqref{J5} would be sufficient to measure $\omega$, $\theta$ and $\phi$.

For further discussion of rotating gravitational wave detectors, see~\cite{Kuwahara:2022dyx, MaganaHernandez:2022ayv} and the references cited therein.

\section{Gravitational Time Delay}

Consider the exterior gravitational field of a rotating astronomical source that remains at rest in the background Minkowski spacetime. This rotating source will function as a gravitational lens in the main part of this section. We treat the gravitational field in accordance with the linear approximation of general relativity (GR). It is possible to express linearized GR in a form that is analogous to electrodynamics~\cite{Einstein}. In this framework, gravitational effects can be described in terms of gravitoelectric and gravitomagnetic fields. The former are familiar from the correspondence of GR with Newtonian gravitation, while the latter are non-Newtonian fields generated by mass current and can be expressed as 
\begin{equation}\label{K}
\mathbf{B}_g = \nabla\,\times \mathbf{A}_g\,,
\end{equation}
where $\mathbf{A}_g$ is the gravitomagnetic vector potential. For the sake of concreteness, we assume that the source is stationary, nearly spherical and rotating slowly with constant total angular momentum
  $\mathbf{J}$. Far from the source, we can write
\begin{equation}\label{L}
\mathbf{A}_g(\mathbf{x}) = \frac{G}{c}\,\frac{\mathbf{J} \times \mathbf{x}}{|\mathbf{x}|^3}\,. 
\end{equation}
Moreover,  the corresponding gravitomagnetic field is given by  
\begin{equation}\label{M}
\mathbf{B}_g(\mathbf{x}) = \frac{G}{c}\,\frac{3\,(\mathbf{J}\cdot \mathbf{x})\,\mathbf{x} - \mathbf{J}\,|\mathbf{x}|^2}{|\mathbf{x}|^5}\,. 
\end{equation}
The gravitomagnetic field of the earth has been measured via the Gravity Probe B (GP-B) experiment~\cite{Francis1, Francis2}. For a recent detailed treatment of gravitomagnetism, see Ref.~\cite{Bini:2021gdb} and the references cited therein. 

Next, imagine the propagation of a null signal from a point $P_1: (ct_1, \mathbf{x}_1)$ to a point $P_2: (ct_2, \mathbf{x}_2)$ in the exterior of the rotating gravitational source under consideration here. The path of the signal is lightlike; therefore, 
\begin{equation}\label{Ma}
 g_{\mu \nu} \,dx^\mu dx^\nu = 0\,.  
\end{equation}
With $g_{\mu \nu} = \eta_{\mu \nu} + h_{\mu \nu}$, Equation~\eqref{Ma} implies
\begin{equation}\label{Mb}
 dt - |d\mathbf{x}| = \frac{1}{2}\,h_{\mu \nu} n^\mu n^\nu dt\,,  
\end{equation}
where $h_{\mu \nu}$ is evaluated along the unperturbed signal, namely, $dx^\mu/dt = n^\mu = (1, \mathbf{n})$. Here, $\mathbf{n}$ is a constant unit vector that indicates the unperturbed spatial direction from $P_1$ to $P_2$. Integrating Equation~\eqref{Mb}, we  find
\begin{equation}\label{N}
 t_2 - t_1 = \frac{1}{c}\,|\mathbf{x}_2 - \mathbf{x}_1| +  \Delta_{ge} + \Delta_{gm}\,,
\end{equation}
where, in our linear approximation scheme, the gravitational time delay in this case is the sum of the Shapiro gravitoelectric time delay $\Delta_{ge}$ plus the gravitomagnetic time delay $\Delta_{gm}$. In this section, we are mainly interested in $\Delta_{gm}$ given by~\cite{GMdelay}
\begin{equation}\label{O}
 \Delta_{gm} = -\frac{2}{c^3}\,\int_{P_1}^{P_2} \mathbf{A}_g \cdot d\mathbf{x}\,.  
\end{equation}
Let us note here that, for the sake of simplicity,  we have defined time delay using coordinate time; however, the observer can simply convert it to its proper time in order to obtain an invariant measurable result.  

To evaluate the line integral of the gravitomagnetic potential~\eqref{L} that occurs in Equation~\eqref{O}, we assume that the origin of spatial coordinates coincides with the center of mass of the rotating source and, furthermore,  $|\mathbf{x}_1|$ and $|\mathbf{x}_2|$ are both $\gg GM/c^2$, where $M$ is the mass of the source. Then, from Equation~\eqref{O}, we find 
\begin{equation}\label{P}
 \Delta_{gm} = -\frac{2G\mathbf{J}\cdot \hat{\mathbf{w}}}{c^4}\,\left(\frac{1}{|\mathbf{x}_1|} + \frac{1}{|\mathbf{x}_2|}\right) \frac{|\hat{\mathbf{x}}_1 \times \hat{\mathbf{x}}_2|}{1+ \hat{\mathbf{x}}_1 \cdot \hat{\mathbf{x}}_2}\,,  
\end{equation}
where
\begin{equation}\label{Q} 
\hat{\mathbf{w}} = \frac{\hat{\mathbf{x}}_1 \times \hat{\mathbf{x}}_2}{|\hat{\mathbf{x}}_1 \times \hat{\mathbf{x}}_2|}\,.
\end{equation}
Clearly, $ \Delta_{gm}$ vanishes if $\mathbf{J}$, $\mathbf{x}_1$ and $\mathbf{x}_2$ are in the same plane or $\mathbf{x}_1$ and $\mathbf{x}_2$ are parallel. 

We now consider the rotating source as a gravitational lens for the propagation of electromagnetic and gravitational rays of radiation. Let $\mathcal{D}$ be the impact parameter of the null signal such that $\mathcal{D} \ll |\mathbf{x}_1|$ and $\mathcal{D} \ll |\mathbf{x}_2|$. Then, it is straightforward to show that Equation~\eqref{P} reduces to
\begin{equation}\label{R}
 \Delta_{gm} \approx -\frac{4G\mathbf{J}\cdot \hat{\mathbf{w}}}{c^4\mathcal{D}}\,.  
\end{equation}
It is important to emphasize that these results follow from the basic assumption that the signal travels along a null trajectory. The resulting time delays are therefore independent of the nature and polarization of the signal.  

Observations of gravitational lensing with black holes may reveal the contribution of gravitomagnetic effect in addition to the Shapiro time delay. Assuming a gravitational lens with an impact parameter of the order of the Einstein radius, namely, $R_E = \mathcal{D}$ and using the relation  $R_E = (R_S\,d)^{1/2}$, where $d$ is a typical lens-source distance and $R_S$ is the Schwarzschild radius of the lens, we can write Equation~\eqref{R} as
\begin{equation}\label{Ra}
\Delta_{gm} \approx -\frac{4G\mathbf{J}\cdot \hat{\mathbf{w}}}{c^4(R_S\,d)^{1/2}}\,.  
\end{equation}
For an extreme Kerr  black hole with a maximum angular momentum of $J = G M^2/ c$, the magnitude of gravitomagnetic time delay simplifies to 
$|\Delta_{gm}| \approx (R_S/c)(R_S/d)^{1/2}|\hat{\mathbf{J}} \cdot \hat{\mathbf{w}}|$. For an extreme Kerr black hole of mass $M \sim 10^6 M_{\odot}$ at a cosmological distance that has been detected through microlensing of a gamma-ray burst (GRB), such as the source in~\cite{Kalantari:2021sqy, Kalantari:2022cpe}, the numerical value of the time delay would be at most
$|\Delta_{gm}| \approx  10^{-3}$ sec. This time delay turns out to be much shorter than the sampling rate of the light curve, which is of the order of $10^{-1}$ sec.  The sampling rate of GRB light curve  by FERMI satellite is larger than the gravitomagnetic time delay; therefore, the latter is not detectable in this data. However, future gamma-ray detectors with high sampling rate should take into account the possible contribution of the gravitomagnetic term.

It is worth mentioning that the time delay due to the passage of a free gravitational wave can be simply evaluated using Equation~\eqref{Mb}. That is, 
\begin{equation}\label{Rb}
\Delta_{gw} = \frac{1}{2} n^i n^j \int_{t_1}^{t_2} h_{ij} (t, \mathbf{x}_1 +  \mathbf{n} (t-t_1))\,dt\,.  
\end{equation}
If the plane gravitational wave propagates along the same direction as the null signal, $\Delta_{gw} = 0$ by tranversality; otherwise, $\Delta_{gw}/(t_2-t_1)$ is generally of the order of magnitude of the average amplitude of the incident radiation. For instance, in the case of an incident wave packet with amplitude
\begin{equation}\label{Rc}
h_{ij} (t, \mathbf{x}) = {\rm Re} \sum_\omega \hat{h}_{ij}(\omega) e^{-i\omega(t-\hat{\mathbf{k}}\cdot \mathbf{x})}\,,  
\end{equation}
where $\hat{\mathbf{k}}$ is fixed and $1-\hat{\mathbf{k}}\cdot \mathbf{n}\ne 0$, we have 
\begin{equation}\label{Rd}
\Delta_{gw} = \frac{n^i n^j}{1-\hat{\mathbf{k}}\cdot \mathbf{n}} {\rm Re} \sum_\omega \hat{h}_{ij}(\omega) \frac{\sin(\omega \sigma)}{\omega} e^{-i\omega(\sigma_0 + \sigma)}\,.  
\end{equation}
Here, 
\begin{equation}\label{Re}
\sigma_0 = t_1 - \hat{\mathbf{k}} \cdot \mathbf{x}_1\,, \qquad \sigma = \frac{1}{2} (1-\hat{\mathbf{k}}\cdot \mathbf{n})(t_2-t_1)\,.  
\end{equation}

\section{Propagation of Polarized Gravitational Waves}

Let us return to Equation~\eqref{E} and note that in the rotating frame of the static observer, $\omega^2 = c^2 k^2$  implies
\begin{equation}\label{S1}
(\omega' + \mathbf{H}'\cdot \boldsymbol{\Omega})^2 = c^2 k'^2\,,  
\end{equation}
where $\mathbf{H}' = \pm\, s \hat{\mathbf{k}}'$ and we define the unit vector  $\hat{\mathbf{k}}':= \mathbf{k}'/|\mathbf{k}'|$.  Working to first order in the helicity-rotation coupling, the dispersion relation for radiation propagating in the rotating frame is thus given by
\begin{equation}\label{S2}
\omega'^2 = c^2 k'^2 \mp 2 c s\mathbf{k}'\cdot \boldsymbol{\Omega}\,.  
\end{equation}

According to the gravitational Larmor theorem~\cite{Mash93}, which is the gravitomagnetic version of Einstein's principle of equivalence, the gravitomagnetic field of a rotating source is locally equivalent to the angular velocity of rotation of a frame of reference such that 
\begin{equation}\label{S3}
\frac{1}{c}\,\mathbf{B}_g = - \boldsymbol{\Omega}\,.  
\end{equation}
In electrodynamics, Larmor's theorem originally established a certain equivalence between physics in a rotating frame and in the presence of a magnetic field~\cite{Larmor}; similarly, Equation~\eqref{S3} expresses a certain equivalence between rotation and gravitomagnetism. Therefore, using Equations~\eqref{S2} and~\eqref{S3}, the local dispersion relation for radiation propagating in the exterior field of a rotating source can be written to first order in $|B_g/(c\omega)| \ll 1$ as
\begin{equation}\label{S4}
\omega^2 = c^2 k^2 \pm 2  s\mathbf{k}\cdot \mathbf{B}_g\,. 
\end{equation}
It follows from this relation that 
\begin{equation}\label{S5}
\omega = c k \pm \frac{1}{c} s \hat{\mathbf{k}}\cdot \mathbf{B}_g\,. 
\end{equation}
We need to extend this local relation to the spacetime of the rotating source. In the WKB limit ($\omega \to \infty$), the radiation propagates along a null geodesic ray that bends in the gravitational field due to the mass $M$ and angular momentum $J$ of the source. The motion of the ray is independent of the polarization  of the radiation. The helicity-rotation-gravity coupling evident in Equation~\eqref{S5} constitutes a weak perturbation that  occurs around the ray; that is, the polarization of the radiation gives rise to a small correction in the motion of the ray.  To simplify matters, we can ignore the bending of the ray and work in a global inertial frame in Minkowski spacetime with the understanding that the deviations we find around a straight null path actually occur about a deflected null geodesic ray. 

Consider the propagation of (electromagnetic or gravitational) radiation in the stationary gravitational field of a uniformly rotating source. The frequency of the radiation remains constant in the stationary background. Hamilton's equations for the dispersion relation~\eqref{S5} with $\omega = \mathcal{H}(\mathbf{x}, \mathbf{k})$ can be expressed as
\begin{equation}\label{S6}
\frac{d \mathbf{x}}{dt} = \frac{\partial \mathcal{H}}{\partial \mathbf{k}} = \mathbf{v}_{gr}\,
\end{equation}
and
\begin{equation}\label{S7}
\frac{d \mathbf{k}}{dt} = -\frac{\partial \mathcal{H}}{\partial \mathbf{x}} = \mp s \nabla (\hat{\mathbf{k}}\cdot \mathbf{B}_g)\,.
\end{equation}
Here, the group velocity $\mathbf{v}_{gr}$ can be computed by differentiating Equation~\eqref{S4} with respect to $\mathbf{k}$ and the result is
\begin{equation}\label{S8}
\mathbf{v}_{gr} = \frac{1}{\omega} (\mathbf{k} \pm s\mathbf{B}_g)\,.
\end{equation}

To solve these equations, we assume that
\begin{equation}\label{S9}
 \omega = \mathcal{H}(\mathbf{x}_{+}, \mathbf{k}_{+}) = \mathcal{H}(\mathbf{x}_{-}, \mathbf{k}_{-})\,, 
\end{equation} 
 where 
\begin{equation}\label{S10}
\mathbf{k}_{\pm} = \mathbf{k}_{0} \pm \boldsymbol{\kappa} (t)\,, \qquad \mathbf{x}_{\pm} = \mathbf{x}_{0} \pm \mathbf{q}(t)\,
\end{equation}
correspond to positive ($+$) and negative ($-$) helicity states of the radiation. We treat $ \boldsymbol{\kappa}(t)$ and $\mathbf{q}(t)$ to linear order in accordance with our perturbation scheme; that is, $\mathbf{k}_{0}$ indicates the wave vector of the average unpolarized null signal. Substituting Equation~\eqref{S10} in Hamilton's equations, we find
\begin{equation}\label{S11}
\frac{d \mathbf{x}_0}{dt} = \frac{\mathbf{k}_0}{\omega}\,, \qquad \frac{d\mathbf{q}}{dt} = \frac{1}{\omega}(\boldsymbol{\kappa} + s\mathbf{B}_g)\, 
\end{equation}
and
\begin{equation}\label{S12}
\frac{d \mathbf{k}_0}{dt} = 0\,, \qquad \frac{d\boldsymbol{\kappa}}{dt} = - s \nabla (\hat{\mathbf{k}}_0\cdot \mathbf{B}_g)\,.
\end{equation}
Furthermore, Equations~\eqref{S5},~\eqref{S9} and~\eqref{S10} imply, after some algebra, that
\begin{equation}\label{S13}
\hat{\mathbf{k}}_0\cdot (\boldsymbol{\kappa} + s\mathbf{B}_g)= 0\,, \qquad \omega = |\mathbf{k}_0|\,. 
\end{equation}

These results imply that $\mathbf{k}_0$ is a constant propagation vector and $d \mathbf{x}_0/dt = \hat{\mathbf{k}}_0$ is the equation of motion of the unperturbed null ray. Moreover, we note that in the exterior of the rotating source
\begin{equation}\label{S14}
\mathbf{B}_g = - \nabla \Psi_g\,, \qquad \Psi_g = \frac{G}{c} \frac{\mathbf{J}\cdot \mathbf{x}}{|\mathbf{x}|^3}\,, 
\end{equation}
where $\Psi_g$ is the \emph{gravitomagnetic scalar potential} such that $\Psi_g(\infty) = 0$.  Hence, in Equation~\eqref{S12}, along the unperturbed ray, 
\begin{equation}\label{S15}
\nabla (\hat{\mathbf{k}}_0\cdot \mathbf{B}_g) = - \nabla \left(\frac{d \mathbf{x}_0}{dt} \cdot \nabla \Psi_g\right) =  \frac{d\mathbf{B}_g}{dt}\,,
\end{equation}
since $\hat{\mathbf{k}}_0$ is constant and $\partial_t \mathbf{B}_g = 0$. It then follows from Equation~\eqref{S12} that
\begin{equation}\label{S16}
\frac{d}{dt} (\boldsymbol{\kappa} + s\mathbf{B}_g)= 0\,, \qquad \boldsymbol{\kappa} + s\mathbf{B}_g := \omega \mathbf{V}_0\,,
\end{equation}
where $\mathbf{V}_0 = d\mathbf{q}/dt$ is a constant vector orthogonal to the unperturbed ray by Equation~\eqref{S13}. It is possible to show that the group speed along the ray is equal to the speed of light,  $\mathbf{v}_{gr} \cdot \hat{\mathbf{k}}_0 = 1$, and in this way the solutions to Hamilton's equations are completely specified. 

An important consequence of these solutions is the differential deflection of polarized radiation~\cite{Ramos:2006sb, Mash93}; that is, 
\begin{equation}\label{S17}
\mathbf{x}_{\pm} = (\hat{\mathbf{k}}_0 \pm \mathbf{V}_0) t +{\rm constant}\,, \qquad  \mathbf{k}_{\pm} = \mathbf{k}_{0} \pm \omega \mathbf{V}_0 \mp s \mathbf{B}_g\,, 
\end{equation}
where $\mathbf{V}_0$ is normal to the direction of the null ray, $\hat{\mathbf{k}}_0 \cdot \mathbf{V}_0 = 0$. In the next  two sections, we employ these results to work out two other consequences of our considerations, namely, the gravitational Faraday rotation of linear polarization and the gravitomagnetic time delay of polarized (electromagnetic or gravitational) waves. The basic idea here follows from Equation~\eqref{S17}, namely, along the ray, the two helicity states have different \emph{phase speeds}~\cite{Mash93}. 

\section{Gravitational Faraday Rotation}

The gravitomagnetic rotation of the plane of linear polarization of electromagnetic waves along the direction of a ray was first pointed out by Skrotskii in 1957 and is known as the Skrotskii effect~\cite{Sk}. It is the gravitational analog of the Faraday rotation in electrodynamics. In general relativity, the amplitude of the electromagnetic wave is parallel propagated along the ray in the eikonal approximation and this leads to the Skrotskii effect in the gravitational field of a rotating source.  It is a general phenomenon valid as well for gravitational waves and can be derived as a consequence of the spin-rotation-gravity coupling.  

Let $\mathbb{L}_1$ and $\mathbb{L}_2$ be the amplitudes of the two independent linear polarization states of the radiation. One can be transformed into the other by a rotation of angle $\pi/(2s)$. The independent circular polarization states are then $\mathbb{C}_{\pm} = (\mathbb{L}_1 \pm i \mathbb{L}_2)/\sqrt{2}$. Here, $\mathbb{C}_{+}$ ($\mathbb{C}_{-}$) refers to positive (negative) helicity amplitude. 

Suppose at some initial event at $ t = t_i$ along the ray characterized by $d\mathbf{x}_{0}/dt = \hat{\mathbf{k}}_0$, the wave amplitude $\psi$ is $\mathbb{L}_1$; that is, $\psi (t_i) = \mathbb{L}_1 = (\mathbb{C}_{+} + \mathbb{C}_{-})/\sqrt{2}$. Then, as the wave propagates along the ray, its amplitude is given by
\begin{equation}\label{R1}
\psi  (t)=  \frac{1}{\sqrt{2}} \left(\mathbb{C}_{+} e^{f_{+}} +  \mathbb{C}_{-} e^{f_{-}}\right)\,,
\end{equation}
where, using the results of the previous section,
\begin{equation}\label{R2}
f_{\pm}  = - i \omega (t-t_i) + i \int_{\bar{\mathbf{x}}_i}^{\bar{\mathbf{x}}} \mathbf{k}_{\pm} \cdot d\mathbf{x}_0\,.
\end{equation}
Here, $\bar{\mathbf{x}}_i = \mathbf{x}_{0}(t_i)$ and $\bar{\mathbf{x}} = \mathbf{x}_{0}(t) = \bar{\mathbf{x}}_i  +  \hat{\mathbf{k}}_0 (t-t_i)$. Along the ray, $\mathbf{V}_0 \cdot \hat{\mathbf{k}}_0 = 0$ and
\begin{equation}\label{R3}
\mathbf{k}_{\pm} \cdot d\mathbf{x}_0 =  (\mathbf{k}_0 \mp s\mathbf{B}_g)\cdot d\mathbf{x}_0\,.
\end{equation}
That is, the different phase speeds have to do with the component of $\mathbf{B}_g$ along the ray. From $\mathbf{k}_0 \cdot d\mathbf{x}_0 = \omega \,dt$, we find
\begin{equation}\label{R4}
f_{\pm}  =  \mp i \Theta\,,
\end{equation}
where 
\begin{equation}\label{R5}
\Theta = s \int_{\bar{\mathbf{x}}_i}^{\bar{\mathbf{x}}} \mathbf{B}_g\cdot d\mathbf{x}_0\,.
\end{equation}
Equations~\eqref{R1} and~\eqref{R4} imply
\begin{equation}\label{R6}
\psi  (t) =  \mathbb{L}_1 \cos \Theta + \mathbb{L}_2 \sin \Theta\,.
\end{equation}
Thus $\Theta$ is the angle of rotation of the linear polarization state along the ray and can be expressed as 
\begin{equation}\label{R7}
\Theta =  - \frac{s}{c^2} [\Psi_g(\bar{\mathbf{x}}) - \Psi_g(\bar{\mathbf{x}}_i)]\,. 
\end{equation}
Alternatively, we can write 
\begin{equation}\label{R8}
\frac{d\Theta}{dt} =  \frac{s}{c} \mathbf{B}_g\cdot \hat{\mathbf{k}}_0\,. 
\end{equation}
If the ray propagates from minus infinity to plus infinity, the net angle of rotation is zero. 

The gravitational Faraday effect, which persists in the WKB limit ($\omega \to \infty$), can also be derived in the case of gravitational radiation from the way the gravitational wave amplitude is propagated along the ray within the framework of the eikonal approximation~\cite{Ramos:2006sb}.  For recent applications of the Skrotskii effect, see~\cite{Chakraborty:2021bsb, Li:2022izh} and the references cited therein.

\section{Delay of Polarized Waves}

Suppose we keep track of the arrival of polarized radiation along the null geodesic ray whose direction is characterized by the unit vector $\hat{\mathbf{k}}_0$. Using Equation~\eqref{S17}, we find that the wave vectors of the two helicity states along this direction are given by 
\begin{equation}\label{D1}
\mathbf{k}_{\pm} \cdot \hat{\mathbf{k}}_0 = \omega \mp  s \hat{\mathbf{k}}_0 \cdot \mathbf{B}_g\,. 
\end{equation}
The different phase speeds along the ray lead to the time delay of polarized radiation. 

Imagine waves of frequency $\omega$ involving both helicity states that start propagating along the ray from $\bar{\mathbf{x}}_i = \mathbf{x}_0(t_i)$ at some initial time $t_i$ and are finally received at $t_f^{\pm}$ by an observer at $\bar{\mathbf{x}}_f = \mathbf{x}_0(t_f)$, where $t_f = (t_f^{+} + t_f^{-})/2$. Here, $t_f^{+}$ ($t_f^{-}$) is the time of reception of positive (negative) helicity wave by the observer at $\bar{\mathbf{x}}_f$. Let us employ Equation~\eqref{D1} and write the phase of polarized radiation along the ray as
\begin{equation}\label{D2}
d\Phi_{\pm} = - \omega dt^{\pm} + \mathbf{k}_{\pm} \cdot d\mathbf{x}_0 = - \omega d(t^{\pm}-t) \mp  s \mathbf{B}_g \cdot d\mathbf{x}_0\,. 
\end{equation}
Then, it follows from the equality of the net variation in the phase of the polarized waves as they propagate from $t_i$ to $t_f^{\pm}$ that
\begin{equation}\label{D3}
\omega (t_f^{\pm} - t_f)=   \mp  s \int_{\bar{\mathbf{x}}_i}^{\bar{\mathbf{x}}_f} \mathbf{B}_g \cdot d\mathbf{x}_0\,. 
\end{equation}
We define the delay of polarized waves along the ray by 
\begin{equation}\label{D4}
\Delta_P :=  t_f^{+} - t_f^{-}\,. 
\end{equation}
Hence,  Equation~\eqref{D3} implies~\cite{Mash93}
\begin{equation}\label{D5}
\Delta_P =  2 \frac{s}{c^2\,\omega}[\Psi_g(\bar{\mathbf{x}}_f) - \Psi_g(\bar{\mathbf{x}}_i)]\,. 
\end{equation}
When the observer is very far away from the gravitational source,  $\Psi_g(\bar{\mathbf{x}}_f)$ may be neglected and we find
\begin{equation}\label{D6}
\Delta_P = - 2 \frac{s\, G}{c^3\omega}\frac{\mathbf{J} \cdot \bar{\mathbf{x}}_i}{|\bar{\mathbf{x}}_i|^3}\,. 
\end{equation}

To illustrate this result for the case of gravitational waves ($s=2$), let us consider the merger of a binary black hole system as the source of the gravitational waves. To simplify matters, we imagine that the result of the merger is a nearly extreme Kerr black hole of mass $M$ and angular momentum $J$ that approaches $GM^2/c$. For the gravitational waves that originate near the horizon of the final black hole, 
$|\bar{\mathbf{x}}_i| \sim 2GM/c^2$ and we find from Equation~\eqref{D6} that at most
\begin{equation}\label{D6a}
|\Delta_P| \sim \frac{1}{\omega}\,. 
\end{equation}
The LIGO and Virgo detectors are sensitive in the frequency range $\nu = \omega/(2\pi)$ of $10$~Hz to $10$~kHz~\cite{ligosens}. The corresponding time delay $|\Delta_P|$ from Equation~\eqref{D6a} would then be in the range of 16 $\mu$sec to 16 msec.  For frequencies  $\nu \sim 100$ Hz, the magnitude of time delay would be at most  $|\Delta_P| \sim1$ msec.  Let us note that $|\Delta_P|$ is linearly proportional to the mass $M$ of the final black hole, since $\omega$ is inversely proportional to $M$; therefore,  the arrival time delay of gravitational waves with different circular polarizations could be larger with more massive binary black hole mergers. The trigger time  of the two detectors in LIGO is about $7$ msec for the event designated as GW150914~\cite{LIGOScientific:2016vbw}. We conclude that the time resolution of future detectors with improved capabilities could possibly resolve arrival time delays of order milliseconds for different circular polarization states of incident gravitational radiation.

\section{Discussion}

The helicity of linearized gravitational waves can couple to the rotation of an observer or the gravitomagnetic field of a rotating astronomical body. We describe the pertinent modifications of the Doppler effect and aberration for the reception of polarized gravitational waves by rotating observers. Moreover, the propagation of polarized gravitational waves in the field of a rotating mass is considered and the corresponding  gravitational Faraday rotation and time delay are studied. The effects of polarized gravitational waves are issues of current interest~\cite{Wu:2019kyb, Oancea:2022szu}.  In the light of the recent discovery of gravitational waves, we provide simple estimates for some of the consequences of the spin-rotation-gravity coupling  involving incident transient gravitational waves. The resulting effects are small and might be of interest for future detectors of gravitational radiation.

\section*{Acknowledgments}
 
B. M. is grateful to Lorenzo Iorio for helpful discussions. 

\appendix


\begin{thebibliography}{99}

\bibitem{MTW}
Misner, C.W.; Thorne, K.S.; Wheeler, J.A.
\emph{Gravitation};
Freeman: San Francisco, USA, 1973.

%\cite{LIGOScientific:2016aoc}
\bibitem{LIGOScientific:2016aoc}
Abbott, B.P.; et al. [LIGO Scientific and Virgo]
Observation of Gravitational Waves from a Binary Black Hole Merger.
\emph{Phys. Rev. Lett.} \textbf{2016}, \textit{116}, no.6, 061102.
%doi:10.1103/PhysRevLett.116.061102
[arXiv:1602.03837 [gr-qc]]
%8372 citations counted in INSPIRE as of 15 Sep 2022

%\cite{Akhshi:2021nsy}
\bibitem{Akhshi:2021nsy}
Akhshi, A.; Alimohammadi, H.; Baghram, S.; Rahvar, S.; Tabar, M.R.R.;  Arfaei, H.
A template-free approach for waveform extraction of gravitational wave events.
\emph{Sci. Rep.} \textbf{2021},  \textit{11}, no.1, 20507.
%doi:10.1038/s41598-021-98821-z
%1 citations counted in INSPIRE as of 15 Sep 2022


\bibitem{AnBr}
Anile, A.M.; Breuer, R.A.
Gravitational Stokes Parameters.
\emph{Astrophys. J.} \textbf{1974}, \textit{189}, 39--50. 

\bibitem{Mash78}
Mashhoon, B.
On Tidal Resonance.
\emph{Astrophys. J.} \textbf{1978}, \textit{223}, 285--298. 

%\cite{Mashhoon:2021qtc}
\bibitem{Mashhoon:2021qtc}
Mashhoon, B.
Gravitomagnetic Stern--Gerlach Force.
\emph{Entropy} \textbf{2021}, \textit{23}, no.4, 445.
%doi:10.3390/e23040445
[arXiv:2102.06433 [gr-qc]]
%3 citations counted in INSPIRE as of 23 Sep 2022

%\cite{Ruggiero:2020oxo}
\bibitem{Ruggiero:2020oxo}
Ruggiero, M.L.; Ortolan, A.
Gravitomagnetic resonance in the field of a gravitational wave.
\emph{Phys. Rev. D} \textbf{2020}, \textit{102}, no.10, 101501.
%doi:10.1103/PhysRevD.102.101501
[arXiv:2011.01663 [gr-qc]]
%13 citations counted in INSPIRE as of 23 Sep 2022

%\cite{Ruggiero:2021qnu}
\bibitem{Ruggiero:2021qnu}
Ruggiero, M.L.
Gravitational waves physics using Fermi coordinates: a new teaching perspective.
\emph{Am. J. Phys.} \textbf{2021}, \textit{89}, 639.
%doi:10.1119/10.0003513
[arXiv:2101.06746 [gr-qc]]
%3 citations counted in INSPIRE as of 23 Sep 2022

%\cite{Ruggiero:2022gzl}
\bibitem{Ruggiero:2022gzl}
Ruggiero, M.L.
Gravitomagnetic induction in the field of a gravitational wave.
\emph{Gen. Relativ. Gravit.} \textbf{2022}, \textit{54}, no.9, 97.
%doi:10.1007/s10714-022-02983-8
[arXiv:2204.08914 [gr-qc]]
%0 citations counted in INSPIRE as of 23 Sep 2022


\bibitem{Brad}
Bradley, J.
A Letter from the Reverend Mr. James Bradley Savilian Professor of Astronomy at Oxford, and F.R.S. to Dr.Edmond Halley Astronom. Reg. $\&$c. Giving an Account of a New Discovered Motion of the Fix'd Stars. 
\emph{Phil. Trans. Roy. Soc. (London)} \textbf{1728}, \textit{35}, 637--661. 


\bibitem{BMr}
Mashhoon, B.
Electrodynamics in a Rotating Frame of Reference.
\emph{Phys. Lett. A} \textbf{1989}, \textit{139}, 103--108.

%\cite{Hauck:2003gy}
\bibitem{Hauck:2003gy}
Hauck, J.C.; Mashhoon, B.
Electromagnetic waves in a rotating frame of reference.
\emph{Ann. Phys. (Berlin)} \textbf{2003},  \textit{12}, 275--288.
%doi:10.1002/andp.200310011
[arXiv:gr-qc/0304069 [gr-qc]]
%16 citations counted in INSPIRE as of 02 Aug 2022

%\cite{Ramos:2006sb}
\bibitem{Ramos:2006sb}
Ramos, J.; Mashhoon, B.
Helicity-rotation-gravity coupling for gravitational waves.
\emph{Phys. Rev. D} \textbf{2006}, \textit{73}, 084003.
%doi:10.1103/PhysRevD.73.084003
[arXiv:gr-qc/0601054 [gr-qc]]
%17 citations counted in INSPIRE as of 02 Aug 2022


\bibitem{BMB}
Mashhoon, B.
\emph{Nonlocal Gravity};
Oxford University Press: Oxford, UK, 2017. 

\bibitem{Danner}
Danner, A.; Demirel, B.; Kersten, W.; Lemmel, H.; Wagner, R.; Sponar, S.; Hasegawa, Y.
Spin-rotation coupling observed in neutron interferometry.
\emph{npj Quantum Information} \textbf{2020}, \textit{6}, 23. 
%https://doi.org/10.1038/s41534-020-0254-8


\bibitem{BMa}
Mashhoon, B.
Electrodynamics in a Linearly Accelerated System.
\emph{Phys. Lett. A} \textbf{1987}, \textit{122}, 67--72.

%\cite{Mashhoon:2000jq}
\bibitem{Mashhoon:2000jq}
Mashhoon, B.
Gravitational couplings of intrinsic spin.
\emph{Classical Quantum Gravity} \textbf{2000},  \textit{17}, 2399--2410.
%doi:10.1088/0264-9381/17/12/312
[arXiv:gr-qc/0003022 [gr-qc]]
%99 citations counted in INSPIRE as of 02 Aug 2022

%\cite{Bini:2004kz}
\bibitem{Bini:2004kz}
Bini, D.; Cherubini, C.; Mashhoon, B.
Spin, acceleration and gravity.
\emph{Classical Quantum Gravity} \textbf{2004}, \textit{21}, 3893--3908.
%doi:10.1088/0264-9381/21/16/005
[arXiv:gr-qc/0406061 [gr-qc]]
%18 citations counted in INSPIRE as of 02 Aug 2022


%\cite{Bini:2004ay}
\bibitem{Bini:2004ay}
Bini, D.; Cherubini, C.; Geralico, A.; Mashhoon, B.
Spinning particles in the vacuum C metric.
\emph{Classical Quantum Gravity} \textbf{2005}, \textit{22}, 709--722.
%doi:10.1088/0264-9381/22/4/005
[arXiv:gr-qc/0411098 [gr-qc]]
%8 citations counted in INSPIRE as of 02 Aug 2022

%\cite{Mashhoon:2013jaa}
\bibitem{Mashhoon:2013jaa}
Mashhoon, B.; Obukhov, Y.N.
Spin Precession in Inertial and Gravitational Fields.
\emph{Phys. Rev. D} \textbf{2013}, \textit{88}, no.6, 064037.
%doi:10.1103/PhysRevD.88.064037
[arXiv:1307.5470 [gr-qc]]
%15 citations counted in INSPIRE as of 02 Aug 2022



\bibitem{GWarray}
Schutz, B.F.  
Networks of gravitational wave detectors and three figures of merit.
\emph{Classical Quantum Gravity} \textbf{2011}, \textit{28}, 12. 



\bibitem{SathyaSchutz:LivingReviews}
Sathyaprakash, B.S.; Schutz, B.F. Physics, astrophysics, and cosmology with
gravitational waves. 
\emph{Living Reviews in Relativity} \textbf{2009}, \textit{12}, 2.
%doi:10.12942/lrr-2009-2
[arXiv:0903.0338 [gr-qc]]




%\cite{Kuwahara:2022dyx}
\bibitem{Kuwahara:2022dyx}
Kuwahara, N.; Asada, H.
Earth rotation and time-domain reconstruction of polarization states for continuous gravitational waves from known pulsars.
\emph{Phys. Rev. D} \textbf{2022}, \textit{106}, no.2, 024051.
%doi:10.1103/PhysRevD.106.024051
[arXiv:2202.00171 [gr-qc]]
%0 citations counted in INSPIRE as of 27 Nov 2022



%\cite{MaganaHernandez:2022ayv}
\bibitem{MaganaHernandez:2022ayv}
Maga\~na Hernandez, I.
Measuring the polarization content of gravitational waves with strongly lensed binary black hole mergers.
[arXiv:2211.01272 [gr-qc]]
%0 citations counted in INSPIRE as of 27 Nov 2022



\bibitem{Einstein} 
Einstein, A.
\emph{The Meaning of Relativity}; 
Princeton University Press: Princeton, NJ, USA, 1955.

\bibitem{Francis1}
Everitt, C.W.F.; DeBra, D.B.; Parkinson, B.W.; Turneaure, J.P.; Conklin, J.W.; Heifetz, M.I.; Keiser, G.M.; Silbergleit, A.S.; Holmes, T.; Kolodziejczak, J.; et al. 
Gravity Probe B: Final results of a space experiment to test general relativity.
\emph{Phys. Rev. Lett.} {\bf 2011}, \emph{106}, 221101.
[arXiv:1105.3456 [gr-qc]]


\bibitem{Francis2}
Everitt, C.W.F.; Muhlfelder, B.; DeBra, D.B.; Parkinson, B.W.; Turneaure, J.P.; Silbergleit, A.S.; et al. 
The Gravity Probe B test of general relativity. 
\emph{Classical Quantum Gravity} {\bf 2015}, \emph{32}, 224001.


%\cite{Bini:2021gdb}
\bibitem{Bini:2021gdb}
Bini, D.; Mashhoon, B.; Obukhov, Y.N.
Gravitomagnetic helicity.
\emph{Phys. Rev. D} \textbf{2022}, \textit{105}, no.6, 064028.
%doi:10.1103/PhysRevD.105.064028
[arXiv:2112.07550 [gr-qc]]
%1 citations counted in INSPIRE as of 07 Aug 2022


\bibitem{GMdelay}
Ciufolini, I.; Kopeikin, S.; Mashhoon, B.; Ricci, F.
On the Gravitomagnetic Time Delay.
\emph{Phys. Lett. A} \textbf{2003}, \textit{308}, 101--109.
[arXiv:gr-qc/0210015]


%\cite{Kalantari:2021sqy}
\bibitem{Kalantari:2021sqy}
Kalantari, Z.; Ibrahim, A.; Tabar, M.R.R.; Rahvar, S.
Imprints of Gravitational Millilensing on the Light Curve of Gamma-Ray Bursts.
\emph{Astrophys. J.} \textbf{2021}, \textit{922}, no.1, 77.
%doi:10.3847/1538-4357/ac1c06
[arXiv:2105.00585 [astro-ph.CO]]
%7 citations counted in INSPIRE as of 03 Aug 2022

%\cite{Kalantari:2022cpe}
\bibitem{Kalantari:2022cpe}
Kalantari, Z.; Rahvar, S.; Ibrahim, A.
Fermi-GBM Observation of GRB 090717034: $\chi^{2}$ Test Confirms Evidence of Gravitational Lensing by a Supermassive Black Hole with a Million Solar Mass.
\emph{Astrophys. J.} \textbf{2022}, \textit{934}, no.2, 106.
%doi:10.3847/1538-4357/ac7da9
[arXiv:2205.05278 [astro-ph.HE]]
%0 citations counted in INSPIRE as of 03 Aug 2022


\bibitem{Mash93}
Mashhoon, B.
On the gravitational analogue of Larmor's theorem.
\emph{Phys. Lett. A} \textbf{1993}, \textit{173}, 347--354.

\bibitem{Larmor}
Larmor, J.
\emph{Aether and Matter};
Cambridge University Press: Cambridge, UK, 1900.

\bibitem{Sk}
Skrotskii, G.V.
The Influence of Gravitation on the Propagation of Light.
\emph{Sov. Phys. Dokl.} \textbf{1957}, \textit{2}, 226--229.


%\cite{Chakraborty:2021bsb}
\bibitem{Chakraborty:2021bsb}
Chakraborty, C.
Gravitational analog of Faraday rotation in the magnetized Kerr and Reissner-Nordstr\"om spacetimes.
\emph{Phys. Rev. D} \textbf{2022}, \textit{105}, no.6, 064072.
%doi:10.1103/PhysRevD.105.064072
[arXiv:2106.03520 [gr-qc]]
%6 citations counted in INSPIRE as of 08 Nov 2022



%\cite{Li:2022izh}
\bibitem{Li:2022izh}
Li, Z.; Qiao, J.; Zhao, W.; Er, X.
Gravitational Faraday Rotation of gravitational waves by a Kerr black hole.
\emph{JCAP} \textbf{2022}, \textit{10}, 095.
%doi:10.1088/1475-7516/2022/10/095
[arXiv:2204.10512 [gr-qc]]
%2 citations counted in INSPIRE as of 27 Nov 2022



\bibitem{ligosens}
Martynov, D.V.; Hall, E.D.; Abbott, B.P.; Abbott, R.; Abbott, T.D.; Adams, C.; Adhikari, R.X.; Anderson, R.A.; Anderson, S.B.; Arai, K.; et al.
The Sensitivity of the Advanced LIGO Detectors at the Beginning of Gravitational Wave Astronomy.
\emph{Phys. Rev. D} \textbf{2016}, \textit{93}, 112004.
[arXiv:1604.00439 [astro-ph.IM]]


%\cite{LIGOScientific:2016vbw}
\bibitem{LIGOScientific:2016vbw}
Abbott, B.P.; et al. [LIGO Scientific and Virgo]
GW150914: First results from the search for binary black hole coalescence with Advanced LIGO.
\emph{Phys. Rev. D} \textbf{2016}, \textit{93}, no.12, 122003.
%doi:10.1103/PhysRevD.93.122003
[arXiv:1602.03839 [gr-qc]]
%359 citations counted in INSPIRE as of 14 Sep 2022 


%\cite{Wu:2019kyb}
\bibitem{Wu:2019kyb}
Wu, Q.; Zhu, W.; Feng, L.
Testing the Wave-Particle Duality of Gravitational Wave Using the Spin-Orbital-Hall Effect of Structured Light.
\emph{Universe} \textbf{2022}, \textit{8}, no.10, 535.
%doi:10.3390/universe8100535
[arXiv:1904.05380 [gr-qc]]
%1 citations counted in INSPIRE as of 26 Nov 2022



%\cite{Oancea:2022szu}
\bibitem{Oancea:2022szu}
Oancea, M.A.; Stiskalek, R.; Zumalac\'arregui, M.
From the gates of the abyss: Frequency- and polarization-dependent lensing of gravitational waves in strong gravitational fields.
[arXiv:2209.06459 [gr-qc]]
%1 citations counted in INSPIRE as of 26 Nov 2022



\end{thebibliography}
\end{document}